\documentclass[aps,prb,amsmath,amssymb,reprint,superscriptaddress,floatfix,showpacs]{revtex4-1}

\usepackage{graphicx}
\usepackage{dcolumn}
\usepackage{bm}

\usepackage{color}

\begin{document}


\title{Atomic and electronic structures of a vacancy in amorphous silicon} 


\author{Andreas \surname{Pedersen}}
\affiliation{Faculty of Physical Sciences and Science Institute, University of Iceland, 107 Reykjav\'{\i}k, Iceland}
\author{Laurent \surname{Pizzagalli}}
\affiliation{Departement of Physics and Mechanics of Materials,Institut P$^{\prime}$, CNRS UPR 3346, Universit\'{e} de Poitiers, SP2MI, Boulevard Marie et Pierre Curie, TSA 41123
, 86073 Poitiers Cedex 9, France}
\email{Laurent.Pizzagalli@univ-poitiers.fr}
\author{Hannes \surname{J{\'o}nsson}}
\affiliation{Faculty of Physical Sciences and Science Institute, University of Iceland, 107 Reykjav\'{\i}k, Iceland}


\date{\today}

\begin{abstract}
Locally, the atomic structure in well annealed amorphous silicon appears similar to that of crystalline silicon. We address here the question whether a point defect, specifically a vacancy, in amorphous silicon also resembles that in the crystal. From density functional theory calculations of a large number of nearly defect free configurations, relaxed after an atom has been removed, we conclude that there is little similarity.  The analysis is based on formation energy, relaxation energy, bond lengths, bond angles, Vorono\"i volume, coordination, atomic charge and electronic gap states. All these quantities span a large and continuous range in amorphous silicon and while the removal of an atom leads to the formation of one to two bond defects and to a lowering of the local atomic density, the relaxation of the bonding network is highly effective, and the signature of the vacancy generally unlike that of a vacancy in the crystal.\end{abstract}


\maketitle 


\section{Introduction} \label{s:intro}

Vacancy-related defects in crystalline silicon (c-Si) 
are
some of the most studied classes of defects 
 because of the major role of silicon in electronic-based applications. The properties of the simplest
defect, 
the mono-vacancy, are now well understood
and it serves as
a standard benchmark for electronic structure calculation methods and interatomic potential functions. A Jahn-Teller distortion
reduces
the natural $T_d$ symmetry to $D_{2d}$, thus breaking the degeneracy of deep levels in the electronic gap~\cite{Pus98PRB}. 
Formation energy of about 3.5--4.4~eV is obtained 
from electronic structure calculations~\cite{Elm04PRB,Cal06PRL,Kal14PRB},
in good agreement with available measurements~\cite{Dan86PRL,Sud16PRB}.

Conversely, little is known 
about
the properties of vacancies in the amorphous phase of silicon (a-Si). When it is prepared by ion implantation, thus free of hydrogen, experimental investigations suggest the existence of point defects that are analogous 
to vacancies in Si-c
~\cite{Hov92PRL}. In similarly prepared materials, Laaziri 
et al.
found a deficit in coordination, which suggests the presence of vacancy-related defects~\cite{Laa99PRL,Laa99PRB}. 
The concentration of these defects 
is expected to depend on the degree of relaxation. The heat released during annealing processes is compatible with a point-defect annihilation mechanism~\cite{Roo89PRL,Roo91PRB}. 

Moreover, there have been some theoretical studies of vacancies in a-Si. In their pioneering work, Kim and co-workers showed using tight-binding calculations that monovacancy creation was not associated with deep gap states,  
conversely to what is observed to occur in Si-c~\cite{Kim99PRB}. Another investigation revealed that removing one atom from a disordered silicon model lead to either the full recovery of the bond network or the creation of a stable vacancy~\cite{Mir04JNCS}. Only four configurations were studied in each of these studies, which questions the general character of these conclusions. A large number of configurations is required in order to obtain good statistics. 

In two recent works clear improvements were made addressing this aspect and a total of 216~\cite{Url08PRB} and 1000 configurations~\cite{Jol13PRB} were studied. 
However, atomic interactions
were in these studies modeled by either
tight-binding~\cite{Url08PRB} or 
empirical potential functions~\cite{Jol13PRB}, which 
are known not to give
an accurate description of defects in covalent materials. Another issue 
relates to
the occurrence of vacancies 
with a negative formation energy, as 
observed
in two previous investigations~\cite{Mir04JNCS,Jol13PRB}. This would suggest that the 
a-Si
model used as a starting configuration is not fully relaxed. There is 
therefore
a need for additional calculations to address the above
mentioned
issues.  

An interesting additional point concerns the identification of vacancies in the 
a-Si
material. In c-Si, a vacancy is well characterized and leaves several structural and electronic  
footprints, which can
be used for identification. The question is whether such a similar picture 
could be valid
in the disordered phase. Miranda
et al.
hinted that it might be difficult to identify vacancies using only structural information~\cite{Mir04JNCS}. This point was later confirmed by Urli and co-workers who proposed a criterion involving atomic charge and volume. In fact, their tight-binding investigations suggested that atoms characterized by an excess of charge and a large volume are likely to be neighbors to a vacancy~\cite{Url08PRB}. It is important to test the generality of this 
vacancy
criterion with 
a more accurate description of the atomic interactions
and other 
a-Si models. 

%

In this article we describe investigations of the properties of vacancies in 
a-Si using density functional theory (DFT)
calculations applied to more than one thousand configurations. 
Five highly
optimized a-Si configurations were considered as initial systems, yielding only positive vacancy formation energies. 
We analyze and discuss
the energetics associated with vacancy creation, as well as changes occurring in structure, connectivity and electronic properties, during relaxation. Finally, we 
discuss
 the existence of efficient markers to identify vacancies, as well as the suitability of previously proposed criteria. 

\section{Models and numerical simulations} \label{s:method}

Calculations were performed in the framework of DFT~\cite{Hoh64PR,Koh65PR}, using the VASP package~\cite{Kre99PRB}. The Perdew-Burke-Ernzerhof functional~\cite{Per96PRL} approximation was used to describe the valence electrons, and the projected augmented wave formalism~\cite{Blo94PRB} to describe the core electrons. A plane-wave cut-off of 18~Ry and a $\Gamma$-point sampling were found to be sufficient to obtain well converged forces and energies. With these parameters, a 
c-Si
lattice constant of 5.468~\AA\ is obtained. We also checked the suitability of our setup by calculating a vacancy in 
c-Si.
After relaxation, we obtained the expected Jahn-Teller distortion, and a formation energy of 3.556~eV in good agreement with available experimental~\cite{Dan86PRL,Sud16PRB} and theoretical~\cite{Elm04PRB,Cal06PRL,Kal14PRB} estimates. 

\begin{table}[t]
\caption{\label{tab:models} Characteristics of several a-Si models, obtained after annealing using DFT/PBE and PDA procedure, starting from a continuous random network (CRN), or from a molecular dynamics liquid-quench using the Stillinger-Weber (SW) or Tersoff (T$_n$) potentials: Number of atoms
($N$),
 number of 3-fold (C$_3$), 4-fold (C$_4$), and 5-fold (C$_5$) coordinated atoms,
density ($\Delta \rho$) and excess energy ($\Delta e$) relative to crystalline silicon.
}
\begin{ruledtabular}
\begin{tabular}{lllllll}
Type 	& $N$ & C$_3$	& C$_4$	& C$_5$	& $\Delta \rho$ & $\Delta e$ [eV/at.]  \\ \hline
CRN 	& 200	& 0	& 200	& 0	& -2.5\% & 0.148  \\
SW		& 214	& 0	& 212	& 2	& -1.3\% & 0.147  \\
T$_1$	& 213	& 0	& 213	& 0	& -0.9\% & 0.148  \\
T$_2$	& 209	& 1	& 207	& 1	& -1.5\% & 0.153  \\
T$_3$	& 213	& 0	& 213	& 0	& -0.7\% & 0.156  \\
\end{tabular}
\end{ruledtabular}
\end{table}

A total of five different models for the initial a-Si configuration were considered. 
One was built from a continuous random model (CRN) procedure~\cite{Bar00PRB}, while the others were 
obtained
 from a liquid quench procedure in which atomic interactions were represented by commonly used empirical potential functions~\cite{Sti85PRB,Ter88PRB}. All 
 systems have been further optimized regarding energy and density 
by applying the point defect annihilation (PDA) method combined with forces and energies from the DFT calculations~\cite{Ped17NJP}, 
leading to high quality 
a-Si
models in excellent agreement with available experiments~\cite{Roo91PRB,Laa99PRB}. Table~\ref{tab:models} summarizes the main characteristics of 
the configurations. 
Even those obtained from the liquid-quench procedure include no or very few coordination defects. A vacancy was created by removing one atom, followed by  
relaxation using conjugated gradients with a force convergence criterion of 0.01~eV/\AA$^{-1}$.
This procedure was carried out for all atoms in all the five configurations, yielding
 1049 vacancy configurations which gives unprecedented statistics. 

\section{Energetics} \label{s:energetics}

\begin{figure}
\includegraphics[width=8.6cm]{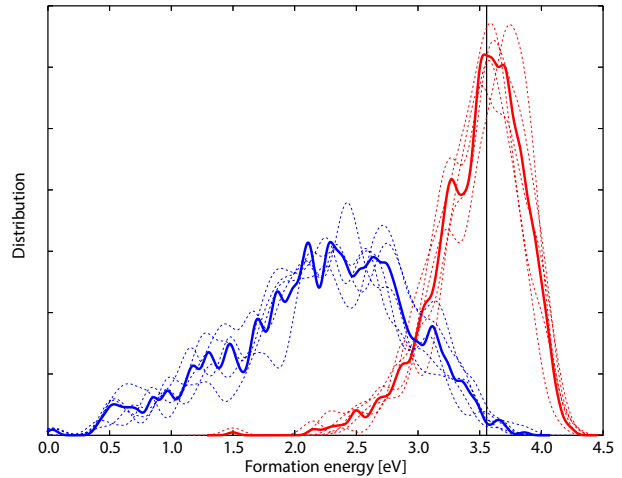}%
\caption{\label{fig:energy}  Distribution of vacancy formation
energies, with ($E_f^r$, blue lines) and without ionic relaxation ($E_f^u$, red lines) 
in amorphous silicon. Dashed thin lines show distributions for each of the five configurations (see Table~\ref{tab:models}), while the thick lines show the average. The vertical 
black
line marks the calculated formation energy for a vacancy in c-Si (3.556~eV).}
\end{figure}

The energy required to create a vacancy in the amorphous material
is defined as 
\begin{equation}
 E_f = E - E_0\times\frac{N-1}{N},
\end{equation}
where $E$ is 
the total energy of the configuration with the vacancy, and $E_0$
is
 the total energy of the initial amorphous configuration
containing
 $N$ 
atoms. Two formation energies can be defined,
$E_f^r$ and $E_f^u$, 
depending whether ionic relaxation is taken into account or not. The vacancy formation energy 
without ionic relaxation $E_f^u$ 
is useful to analyze the effect of ionic relaxation in the amorphous network. The effect of ionic relaxation is defined as the positive quantity 
\begin{equation}
 \Delta E = E_f^u-E_f^r.
\end{equation}

Fig.~\ref{fig:energy} shows the distributions of 
$E_f^r$ and $E_f^u$ 
for all amorphous configurations. Continuous curves were obtained by Gaussian smoothing of width 0.02~eV 
($\sigma$) 
for each system, and 0.001~eV for the summed data. 
The distribution of $E_f^u$ 
and $E_f^r$ is similar for all the configurations. Although 
the
amorphous configurations were initially built using different techniques (CRN or liquid quench), the 
PDA optimization
procedure, which was subsequently applied, yields similar high quality and homogeneous
a-Si
configurations. 
The distribution of $E_f^u$
  can be represented by an asymmetric Gaussian ranging from 2~eV to 4.3~eV approximately, and centered on about 3.7~eV, similar to the formation energy of a relaxed vacancy in c-Si.
  

The $E_f^r$ distribution also exhibits an asymmetric gaussian shape, and is wider and more flat than for 
$E_f^u$
(see fig.~\ref{fig:energy}). 
The most probable value for the formation energy of a vacancy in 
a-Si
ranges from 2.1~eV to 2.7~eV. Assuming a low migration energy, as is the case for a vacancy in c-Si, these values compare fairly well with recent self-diffusion measurements in a-Si~\cite{Kir18PRL}. Another interesting aspect of the distribution is its continuity from slightly above 0~eV to almost 4~eV. 
Since the distribution does not have clear subpeaks, it cannot be used to identify different types of vacancies.

It is interesting to see how low the vacancy formation energy can be (see Fig.~\ref{fig:energy}). This is in contrast to the results of the tight-binding calculations of Kim et al.~\cite{Kim99PRB} where a minimum value of 1.55 eV was found.

The cause of this difference is not clear, but the different level of electronic structure approximation is a possible explanation. Miranda and co-workers reported a negative
vacancy 
formation energy for one case~\cite{Mir04JNCS}. Instead, in our work, we found that all 
vacancies are
associated with positive formation energy, a confirmation that our initial amorphous configurations were fully relaxed. Note that the small bump located close to the 0~eV formation energy in the Fig.~\ref{fig:energy} is associated with a positive value. 

We also attempted to identify a correlation between the unrelaxed formation energy
$E_f^u$
 and the ionic relaxation $\Delta E$, but without success. 
Configurations with large $E_f^u$ apparently do not preferably correspond to large relaxation energy.


\section{Structural relaxation} \label{s:structure}

In this section, we describe 
how
the disordered network 
responds when a vacancy is formed.
The effect of structural relaxation is analyzed through the variation of several indicators such as 
bond
lengths and angles, and Vorono\"i volume. We found that in all cases, the largest changes upon vacancy creation are associated with the 
first-shell 
neighbors of the atom which is removed to create a vacancy. Thus network relaxation mostly occurs in the close vicinity of the created vacancies, and 
only small long range 
relaxation of the network is observed. Therefore, in the following we only examine the variation of structural indicators associated with these first-neighbor atoms, defined as atoms within a distance of 2.8~\AA.
 This separation corresponds to a well defined minimum in the radial distribution function of the amorphous network, between the 
first- and second-shell 
 peaks~\cite{Ped17NJP}. Similar definitions were chosen in previous investigations~\cite{Ber06PRB,Jol13PRB}.

\begin{figure}
\includegraphics[width=8.6cm]{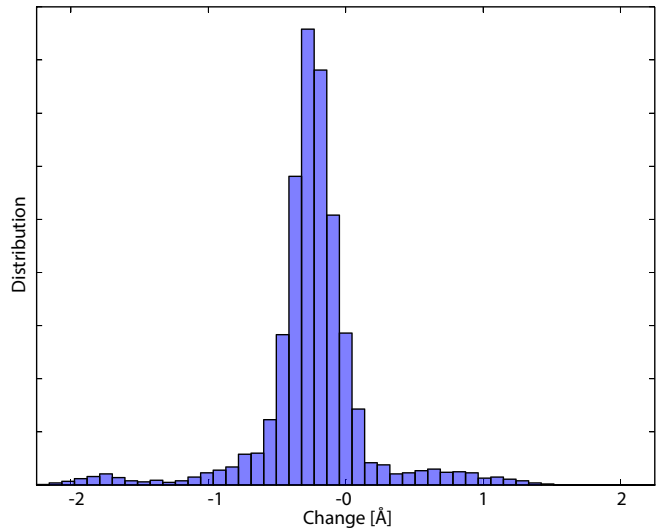}%
\caption{\label{fig:bonddiff} Distribution of the 
change
in the distance 
between the site of the atom removed 
and its 
first-shell neighbors, after vacancy creation.
 Positive (negative) values indicate an outward (inward) relaxation of the atoms surrounding the vacancy.}%
\end{figure}

Fig.~\ref{fig:bonddiff} shows how the first-neighbors of the removed atom are displaced during the relaxation. Most of these atoms are displaced by 0.2-0.3~\AA\ in the direction 
towards 
the newly created vacancy, indicative of an inward relaxation. This is comparable to what occurs in 
c-Si in similar calculations~\cite{Len03JPCM}. An outward relaxation is also obtained in a small proportion of cases. 

\begin{figure}
\includegraphics[width=8.6cm]{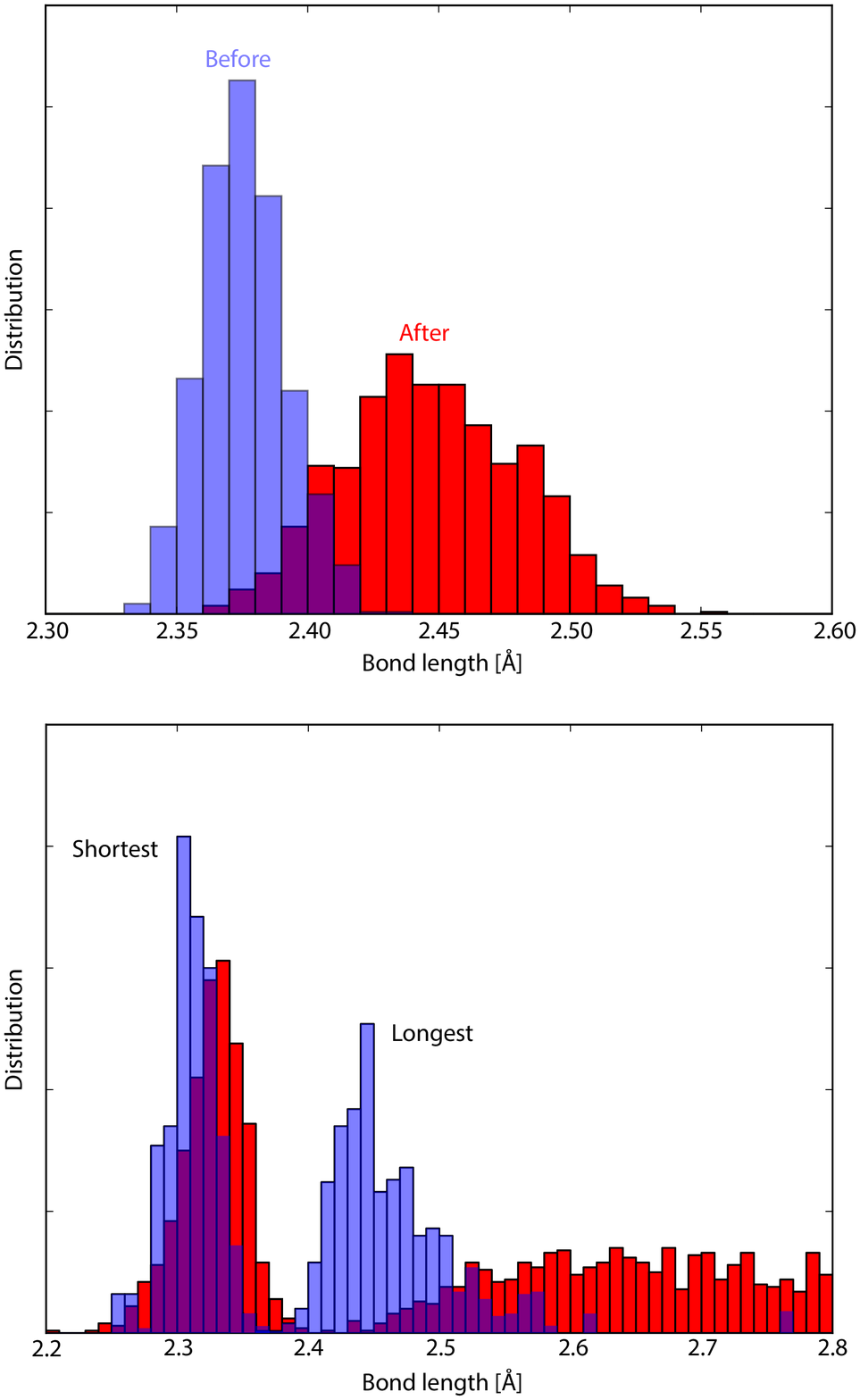}%
\caption{\label{fig:bondlength} Top: Average length distribution of the bonds associated to 
first-shell neighbors of the removed atom (vacancy), before (blue) and after (red) vacancy creation (excluding bonds connecting the removed atom).
Bottom: Longest and shortest bonds distributions.}%
\end{figure}

Additional information is gained by investigating how the length of bonds between first-neighbors
changes
 upon vacancy creation (excluding bonds between these atoms and the atom which is removed). Before vacancy creation, the bond length distribution has a well defined Gaussian-like shape, centered at 2.37~\AA, as shown in the Fig.~\ref{fig:bondlength}. This value is close to the 
 first-shell neighbors
 distance in 
  c-Si~\cite{Laa99PRL,Laa99PRB}. The full width at half maximum (FWHM) is approximately 0.4~\AA. The creation of the vacancy induces a 
 sizable lengthening and the most probable value becomes
   2.43-2.44~\AA. The most frequent response of the amorphous network to vacancy formation
appears to be partial filling of 
the empty space left by the removed atom. The Gaussian shape of the distribution is preserved, but the FWHM is now 0.8~\AA, with extreme values ranging from 2.36 to 2.55~\AA. This illustrates the diversity in the geometry of final configurations. 

In order to analyze the network relaxation
in greater detail, 
 we also examined the changes in the shortest and longest bonds between 
first-neighbors (figure~\ref{fig:bondlength}, bottom). Initially, most of the shortest bonds have a length of 2.3~\AA, with a quite narrow distribution. 
 The formation of a vacancy only
 leads to small changes, with a minor increase of the most probable length to 2.32-2.33~\AA, and an almost negligible widening of the distribution. A different situation is found for the longest bonds. Initially, a Gaussian-like distribution centered on lengths of 2.44-2.45~\AA\ and a FWHM of about 0.7~\AA\ is obtained. After vacancy creation, the distribution changes shape and becomes almost flat. All bond lengths ranging from 2.5~\AA\ to the largest considered value of 2.8~\AA\ are possible and equally likely.
This indicates  
  that the main relaxation of the network involves these long bonds, which are weak and easy to stretch. 
 Conversely, 
the 
 shortest bonds are only slightly lengthened. 

\begin{figure}
\includegraphics[width=8.6cm]{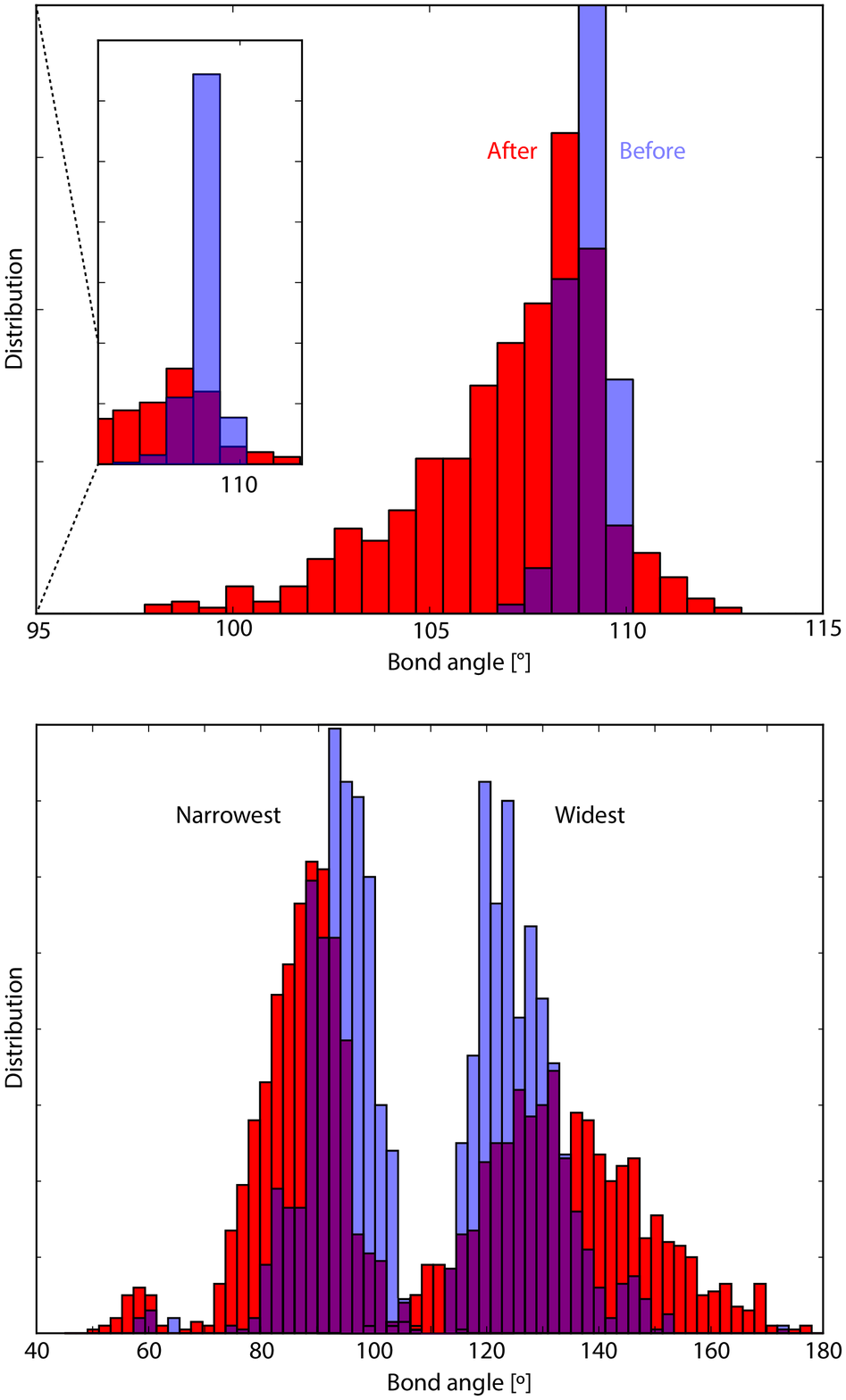}%
\caption{\label{fig:bondangle} Top: Average angle distribution of the bonds associated to 
first-shell neighbors of the removed atom (vacancy), before (blue) and after (red) vacancy creation (excluding bonds connecting the removed atom). The inset graph is an enlarged view showing the full height of the distribution. 
 Bottom: Distributions for narrowest and widest angles.}%
\end{figure}

Changes in angles made between bonds may also provide useful insights about network relaxation. Fig.~\ref{fig:bondangle} shows the distributions of bond angles before and after vacancy creation for 
first-shell neighbors of the removed atom. 
The distribution in the original configurations 
is characterized by a well defined maximum at about 108.2$^\circ$, a value in excellent agreement with experimental determinations~\cite{Laa99PRB,For89PRB} and indicative of extensive annealing~\cite{Ped17NJP}. In 
the vacancy configurations,
the most probable value of the angle between bonds formed by first-neighbors is slightly lowered, by about 0.5$^\circ$. But the most striking difference comes from the width of the distribution. In fact, while bond angles range between 106$^\circ$ and 110$^\circ$ 
in the original configurations,
a much wider range from 98$^\circ$ to 112$^\circ$ is obtained
in the presence of a vacancy, after relaxation.
Overall, it seems that the average bond angle for first-neighbors tends to become more acute, which can be explained by the inward relaxation previously described.  

As for bond lengths, we 
examined the variations of the widest and narrowest bond angles corresponding to vacancy first-neighbors (Fig.~\ref{fig:bondangle}). The narrowest angles distribution is initially centered at 92$^\circ$. 
When the system has relaxed, 
the maximum corresponds to a value of 88$^\circ$, with a minor widening of the distribution. In a few cases, angles as low as 50-60$^\circ$ can be obtained. The influence of the 
relaxation 
is stronger for the widest bond angles. In fact, the initial most probable value of about 123$^\circ$ increases to about 135$^\circ$ 
for the relaxed systems,
associated with a 
sizable 
widening of the distribution. Final bond angles were found to range from about 100$^\circ$ to almost 180$^\circ$. This shows the great diversity of relaxation mechanisms upon vacancy creation in the disordered network. 

\begin{figure}
\includegraphics[width=8.6cm]{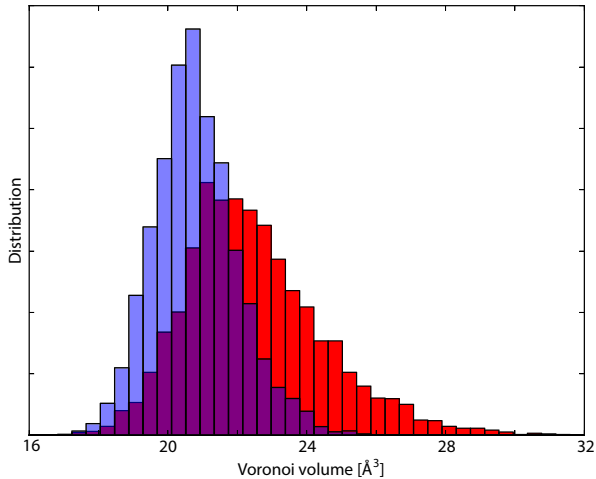}%
\caption{\label{fig:voronoi} Vorono\"i volume distribution associated with vacancy first-neighbors, before (blue) and after (red) vacancy creation.}%
\end{figure}

The atomic volume is another geometrical quantity which
 can be
 used to analyze the network relaxation. Although alternative volume definitions are possible~\cite{Kim99PRB,Mir04JNCS}, we considered here the Vorono\"i volume as in previous investigations~\cite{Url08PRB}. Distributions for vacancy 
first-shell 
  neighbors with and without vacancy are represented in Fig.~\ref{fig:voronoi}. Prior to vacancy formation, a symmetric Gaussian-like distribution is obtained, with a maximum occurring at about 20.6~\AA$^3$ and a FWHM of about 3~\AA$^3$. This most frequent volume corresponds to amorphous silicon which is approximately 1\% less dense than the crystal. After vacancy formation, the distribution is shifted 
 towards
 larger volume with a maximum at 22~\AA$^3$, becomes wider (FWHM of 5~\AA$^3$) and asymmetric due to an increasing weight of large values. This result is
expected, since the decomposition of space according to Vorono\"i rules is complete and the available volume at the vacancy location is essentially shared by the first-neighbors.  

\section{Network connectivity} \label{s:connect}

\begin{table}[t]
\caption{\label{tab:connect} List of changes in the number of bonds (defined as pair-distance within 2.8~\AA) after a vacancy is created, with their occurrence probability for the five a-Si configurations summarized in Table~\ref{tab:models}.
C$_n$ columns report the number of $n$-fold atoms appearing/disappearing during relaxation.}
\begin{ruledtabular}
\begin{tabular}{ccccccc}
C$_2$ & C$_3$  & C$_4$  & C$_5$  & C$_6$  & All & CRN   \\ \hline
0 	& -2	& +2	& 0	& 0	& 43.756\%	&  50.5\% \\
0	& -4	& +4	& 0	& 0	& 24.880\%	&  29.0\% \\
0	& -3	& +2	& +1	& 0	& 15.062\%	&  09.5\% \\
0	& 0	& 0	& 0	& 0	& 04.671\%	&  05.0\% \\
0	& -4	& +2	& +2	& 0	& 02.860\%	&  03.5\% \\
0	& -1	& +2	& -1	& 0	& 02.764\%	&   \\
0	& -3	& +4	& -1	& 0	& 02.574\%	&   \\
0	& -1	& 0	& +1	& 0	& 00.763\%	&  01.5\%  \\
0	& -3	& 0	& +3	& 0	& 00.477\%	&  01.0\%  \\
0	& -2	& 0	& +2	& 0	& 00.381\%	&    \\
0	& -4	& 0	& +4	& 0	& 00.381\%	&    \\
0	& -4	& +3	& 0	& +1	& 00.286\%	&    \\
0	& +1	& 0	& -1	& 0	& 00.286\%	&    \\
0	& -2	& +4	& -2	& 0	& 00.191\%	&    \\
-1	& -2	& +3	& 0	& 0	& 00.191\%	&    \\
0	& -5	& +4	& +1	& 0	& 00.191\%	&    \\
-1	& -3	& +3	& +1	& 0	& 00.095\%	&    \\
0	& -1	& -2	& +3	& 0	& 00.095\%	&    \\
0	& -4	& +6	& -2	& 0	& 00.095\%	&    \\
\end{tabular}
\end{ruledtabular}
\end{table}

\begin{figure}
\includegraphics[width=8.6cm]{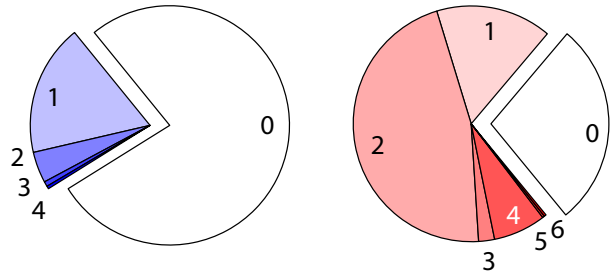}%
\caption{\label{fig:coordination} Proportion of configurations, depending on the number of over-coordinated C$_{5}$ (left pie-chart) and under-coordinated C$_{3}$ (right pie-chart) atoms resulting from the creation of a vacancy followed by relaxation, compared to initial configurations. For instance, in most of cases, no additional C$_{5}$ atoms are created (left, white slice '0'). Also, the most frequent result is the formation of two C$_{3}$ atoms (right, light red slice '2').  
}%
\end{figure}

An important aspect of amorphous systems is the connectivity of the bond network. In the case of silicon, the network is formed by 4-fold coordinated atoms, with a certain proportion of bond defects which depends on the preparation 
method.
The creation of vacancies is obviously expected to result in new bond defects in the network. It is 
therefore interesting to determine to what extent the introduced bond
 defects can be cured during the relaxation. We determined coordination changes for all atoms in the system, using the same distance criterion 
 as
 before, i.e. 2.8~\AA. Note that a more rigorous determination of atomic coordination would require the analysis of the electronic structure in the vicinity of the vacancy. Unfortunately, this is difficult to automate and would require excessive effort for the large
 number of configurations considered here.

The original a-Si configurations include no or very few bond defects. Then, the creation of a vacancy by removing an atom mainly results in 4 fourfold coordinated atoms becoming threefold coordinated, i.e. 4 C$_4$ $\rightarrow$ 4 C$_3$, using the notation in Table~\ref{tab:models}.
We first compare changes in connectivity in vacancy configuration
before and after relaxation. 
Table~\ref{tab:connect} report the observed 
changes,
sorted according to their rate of occurrences. The most frequent
 (about 44\%) corresponds to 2 C$_3$ converted to 2 C$_4$, i.e. the formation of a bond between two atoms 
 that are first-shell neighbors 
  of the vacancy. This is a partial recovery of the network, leaving only 2 C$_3$ atoms. The second most frequent event, accounting for one quarter of the total, is surprisingly a full recovery of fourfold coordination. Then all 4 C$_3$ manage to reform bonds. The third most common event is the conversion of 3 C$_3$ atoms to 2 C$_4$ and 1 C$_5$, leaving two bond defects, a dangling bond but also a weak bond. Obtaining 
  over-coordination
 by removing one atom is counter-intuitive at first, and occurs in about 15\% of the cases. Finally, the fourth most common event with an occurrence probability of about 5\% corresponds to no connectivity modifications during relaxation. This is the 
 configuration closest 
 to a vacancy in 
c-Si,
 with four stable C$_3$ atoms. It is surprising how low the probability of this configuration is.

Several other possible situations with low probabilities have also been obtained. One of the most impressive 
cases
 is the transformation of the 4 C$_3$ atoms to 4 C$_5$ atoms. This was observed four times, and shows that the formation of several weak bond defects can occur
when an atom is removed. 
More generally, we found formation of one or several C$_5$ atoms in about 20\% of the cases. The creation of 
over-coordinated
 atoms 
as a result of introducing a vacancy  is quite surprising. It shows the great diversity of possible network mechanisms for minimizing energy, and is in agreement with previous investigations showing that the creation of bond defects can help lower the energy of amorphous silicon~\cite{Ber06PRB}. Most of the rare 
cases  
   in table~\ref{tab:connect} are related to bond defects already present in the a-Si configurations. This is obviously the 
situation   
   for configurations in which the number of C$_5$ or C$_2$ atoms decreases. For instance, it is interesting to see that the two C$_5$ atoms originally present in the SW a-Si model (see table~\ref{tab:models}) can be annihilated, either with a partial (formation of 4 C$_4$ atoms plus 2 C$_3$ atoms remaining) or a full recovery (formation of 6 C$_4$ atoms) of the 4-fold coordination of the network, although their original separation is about 10~\AA. This shows that in a few cases even the use of local relaxation such as  
conjugate gradient energy minimization 
can lead to network rearrangements on a larger scale. 

To show that the presence of bond defects 
initially
 present in the model has little influence on the statistics, we also report in table~\ref{tab:connect} the results corresponding to the CRN model. One can see that the most frequent relaxation events 
 are the same 
for all models, with similar occurrence probabilities.   

Finally, Fig.~\ref{fig:coordination} shows the proportion of C$_3$ and C$_5$ atoms after vacancy creation and relaxation, compared to the initial a-Si configurations. 
It is clear that in the majority of the cases, 
over-coordinated
 C$_5$ atoms do not form. 
Regarding 
under-coordinated 
  C$_3$ atoms, i.e. dangling bonds, the situation is
more complex. The most likely outcome is the creation of two dangling bonds, followed by the full recovery of the network (no created C$_3$). An examination of the pie chart in Fig.~\ref{fig:coordination} reveals that
the formation of several  
 dangling bonds is quite unlikely. An even number of defects is favored over an odd one, since the formation or breaking of a single bond necessarily involves two atoms.  


\section{Electronic structure} \label{s:electronic}

\begin{table}[t]
\caption{\label{tab:states} Number of cases where none or one to four electronic states appear in the gap when an atom is removed and the configuration relaxed.}
\begin{ruledtabular}
\begin{tabular}{llllll}
Nr of gap states & 0  & 1 & 2 & 3 & 4   \\ \hline
Nr of events & 75  & 318  & 484 & 164  & 8 \\
Percentages  & 7\% & 30\% & 46\% & 16\% & 1\% 
\end{tabular}
\end{ruledtabular}
\end{table}

\begin{figure}
\includegraphics[width=8.6cm]{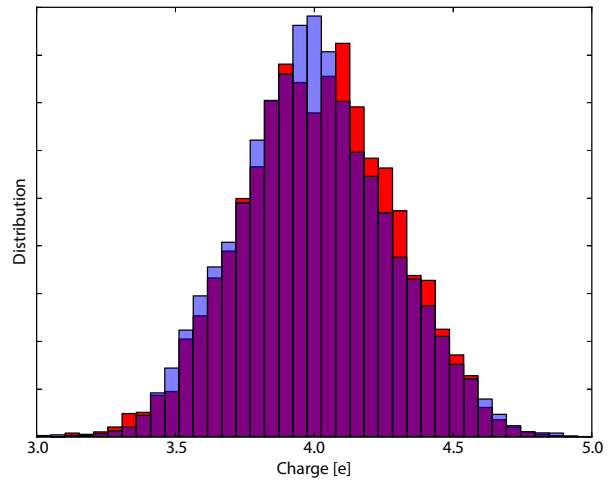}%
\caption{\label{fig:charge} 
Charges for atoms, which are first-shell neighbors of the vacancy, 
computed using a Bader analysis, before (blue) and after (red) vacancy creation.}%
\end{figure}

DFT calculations make it possible to estimate the electronic structure of the material. 
Electronic states, located in the 
band
gap of a-Si are typically associated with dangling or weak bonds. We first investigate the relation between the formation of a vacancy in the amorphous network and the 
appearance 
of states in the gap. Compared to what can occur in c-Si, it is 
difficult to unambiguously discriminate between states corresponding
 to regions with structural defects 
 and ``normal'' states of the amorphous material, even by analyzing the spatial localization of all states characterized by energies in the vicinity of the gap. Here, because of the large number of vacancy configurations, we simply analyzed the number of states appearing in the 
 band
 gap upon vacancy formation. We define such states to have an orbital energy at least 0.25~eV above the valence band edge and 0.25~eV below the conduction band edge.
 Table~\ref{tab:states}
  reports the number of vacancy configurations as a function of the number of new gap states. In most
cases, the creation of
a  vacancy leads to 1 or 2 states appearing in the gap. Less frequent situations are 3 gap states, and no gap states at all. Finally, formation of vacancy with four gap states appears to be rather unlikely, about 1\% of the cases. Overall, we can state that the formation of 
a
vacancy in a-Si is usually associated with 
the generation of 
states located in the 
band
gap. 

We also examined how the creation of a vacancy influences the charge on 
the atoms belonging to the first-shell of neighbors of the vacancy.
The charge distributions, computed following a Bader charge analysis~\cite{Bad90OUP,Hen06CMS}, are shown in Fig.~\ref{fig:charge}. 
Before a 
 vacancy is present in the system, a Gaussian-like distribution is obtained with the expected maximum at 4~e$^-$. Surprisingly, creating a vacancy leads to little changes in the distribution. 
However, the
maximum is now less well defined, with a seemingly double peak structure at 3.9 and 4.1~e$^-$, and there is now a minor asymmetry towards an excess of charge. Nevertheless, the overall initial gaussian structure is preserved. 

\section{Discussion} \label{s:discussion}

The analysis of our data, extracted from our large set of configurations, confirms that it is not possible to firmly identify a vacancy in
 a-Si
  on the basis of structural quantities. We especially studied the structural changes for atoms in 
the closest
 vicinity of the created vacancy, where 
 the changes are most pronounced.
Most of the time, we found an inward relaxation, with small changes in short bonds and a significant stretching of the longest bonds. This is in agreement with Vorono\"i volume calculations, which indicate an increase of the volume for 
these
 atoms. Besides, the analysis of bond angles revealed an increase of the angular dispersion upon vacancy creation. Considering all these indications, it would be tempting to conclude that a cluster of atoms characterized by larger Vorono\"i volumes, and higher disorder in bond angles, should contain a vacancy. Unfortunately, it is not
that  
  simple since we also find many configurations for which the formation of a vacancy is not associated with volume increase or a larger angular disorder. Thus our investigations confirm that a vacancy in a-Si cannot be identified on the sole basis of structural properties.  

\begin{figure}
\includegraphics[width=8.6cm]{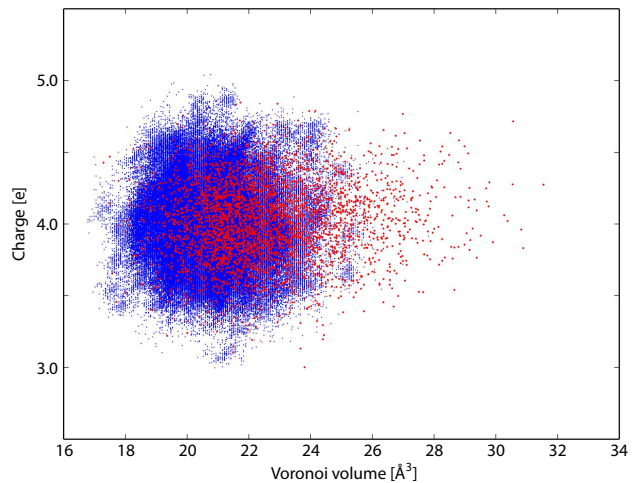}%
\caption{\label{fig:Urlicrit} Charge-volume correlation for all investigated relaxed vacancy configurations. Red dots represent data for atoms 
first-shell neighbors
 of the vacancy while 
blue 
 dots 
represent 
the remaining atoms.}%
\end{figure}

We also investigated charge variations, as well as the number of states appearing in the gap. The former relates to a characterization proposed by Urli and co-workers
involving correlation between volume and charges on atoms in the vicinity of a vacancy~\cite{Url08PRB}. An increase in volume is supposedly correlated with a charge increase of about 0.15~e$^-$ 
 on
 average.  This would lead to
two sets of atoms with distinct properties. 
Our results, however, do not show such a clear correlation.
  We find no significant differences in charge on atoms neighboring the vacancy. The charge-column plot shown in the Fig.~\ref{fig:Urlicrit} exhibits no correlation. While we have no definitive explanation for this difference from the results of Urli et al.~\cite{Url08PRB}, it is likely because of the different level of electronic structure theory employed.
Regarding 
states in the band gap,    
   it seems that vacancy formation is usually (but not always) associated with the apparition of states in the gap. However, since there were already a few states present in the gap for our initial vacancy-free configurations, it is not possible to draw definitive conclusions on this 
issue. 

The
 most striking outcome from our investigations is the changes in coordination during the relaxation of the created vacancy. Although 
 one should be 
 aware that a coordination criterion based only on distance between atoms has some limitations, it appears that a great diversity of final configurations can be obtained. Those cover all conceivable situations, from the creation of several 
 over-coordinated defects to the stabilization of the four 3-fold coordinated atoms as for the
c-Si vacancy. In addition, in about one quarter of the investigated cases, the full recovery
 to a 
 network with no bond defects was obtained. This should placed in context with the distribution of formation energies or relaxation energies, which also span a large range of values. Attempts to find a correlation between these energies and the network relaxation or the final connectivity were unsuccessful. In fact it is possible to find cases for which a low formation energy is associated with a configuration similar to the vacancy in c-Si, and others for which the full recovery of the network is obtained at the cost of an energy in the upper range of the computed values. 

Our investigations show that most of the computed properties change after the creation of a vacancy in the amorphous configuration. But the distribution of possible values
for 
 the final configuration is continuous and homogeneous in all cases, and it always overlaps with the distribution from vacancy-free systems. In fact we did not find a quantity for which there are two well defined and separated sets
 of
  values. In other words, none of the quantities we investigated can be used to define a criterion for vacancy identification. Based on this statement, the obvious question is whether one can consider that a vacancy in amorphous silicon really exist. By analogy with the crystalline phase, it is tempting to consider that a vacancy is located at the center of 4 C$_3$ atoms. Using this definition and a similar criterion for connectivity, Joly and co-workers found that only 45\% of the created vacancies survived the initial structural relaxation~\cite{Jol13PRB}. In our work, using the same vacancy definition, only 5\% were found to be stable following the structural relaxation, the difference being probably due to the superior ability of first-principles calculations to break/create bonds compared to empirical potentials. Joly
et al. also performed long-range relaxation using the activation-relaxation technique, during which almost all remaining vacancies disappeared. Even if such calculations are out of the scope of the present work, it is very likely that a similar result would be obtained for our remaining 5\% configurations. The 
conclusion is therefore that
vacancies similar to those of the crystal may exist in 
a-Si,
but solely as rare and temporary excited states. Finally our investigations on a large set of configurations suggest that removing one atom in an amorphous system can lead to two cases. The first one is the full recovery of the network, thus with no coordination defects, with a locally lower density and probably at the expense of a larger local stress. The second one is the formation of one to two coordination defects, which could be dangling bonds (attached to C$_3$ atoms) or weak bonds (associated with C$_5$ atoms). Dangling bonds are favored over weak bonds, mainly because here we remove atoms from the network. It is likely that a procedure based on the insertion of additional atoms in the network would lead to 
the opposite
 case.  

\section{Conclusion} \label{s:conclusion}

The properties of created vacancies in amorphous silicon have been investigated using DFT. A large set of more than a thousand configurations has been considered in this work, allowing for unprecedented statistics. Highly optimized a-Si models were considered as initial systems, yielding only positive vacancy formation energies. We determined and analyzed the distributions of formation and relaxation energies, bond lengths and angles, Vorono\"i volumes, coordination, atomic charge and creation of gap states. We could not identify markers characteristic of a created vacancy, since possible values for all these quantities span over a large and continuous range. Configurations similar to the vacancy in crystalline silicon appear to be unlikely, except as temporary excited states. Finally, the most probable formed defects in
a-Si
 following the removal of one atom are either a lowering of local density as the bond network fully recovers, or the formation of one to two bond defects. It would be interesting to investigate the stability of these defects by performing long-range relaxation dynamics, using for example the adaptive kinetic Monte Carlo method for long time scale simulations~\cite{Hen01JCP}.  

\section*{Acknowledgments}
We would like to thank G. Barkema for providing us the initial CRN model. The computations were carried out using the Nordic High Performance Computing (NHPC) facility in Iceland. We acknowledge financial support from the Icelandic Research Fund and the French foreign ministry, within the Icelandic-French project call ``Jules Verne''. A. Pedersen also acknowledges the financial support from the R\'egion Poitou-Charentes for his stay at the Pprime Institute. 

%


\end{document}